\documentclass[%
 reprint,
superscriptaddress,
 amsmath,amssymb,
 aps,
pra,
floatfix,
dvipsnames,
]{revtex4-2}

\usepackage[dvipsnames]{xcolor}

\usepackage{color,soul}
\usepackage[normalem]{ulem}

\usepackage{graphicx}
\usepackage{dcolumn}
\usepackage{bm}
\usepackage[hidelinks, breaklinks]{hyperref}

\newcommand\nmberthis{\addtocounter{equation}{1}\tag{\theequation}}
\usepackage{kxbase}
\usepackage{ragged2e}

\begin{document}


\title{Mechanism behind creating qubit gates expressed as interfering quantum pathway amplitudes}

\author{Michael Kasprzak}
\email{mk7592@princeton.edu}
\author{Gaurav Bhole}
\author{Herschel Rabitz}
\affiliation{Princeton University}

\date{\today}

\begin{abstract}
Hamiltonian encoding was introduced as a technique for revealing the mechanism of controlled quantum systems. It does so by decomposing the evolution into pathways between the computational basis states, where each pathway has an associated complex amplitude. The magnitude of a pathway amplitude determines its significance and many pathways constructively and/or destructively interfere to produce the final evolution of the system. In this paper, we apply Hamiltonian encoding to reveal the mechanism behind creating qubit gates implemented via optimal control pulses. An X gate, two CNOT gates, and a SWAP gate are examined to determine the degree of interference involved and to demonstrate that different optimal controls produce distinct mechanisms. Although the detailed mechanism for creating any gate depends on the nature of the control field, the mechanism analysis tools are generic. The presented gates and their mechanisms in this paper are thus illustrative and a researcher may apply these same tools to any gate with a suitable optimal control field. 
\end{abstract}

\maketitle

\section{Introduction}
Creating a viable platform for quantum computing is an area of great interest and potential. Many different approaches have been suggested and each comes with different strategies for creating qubits and implementing quantum gates on the system \cite{10.1063/1.5088164, PhysRevA.57.120, RevModPhys.79.135, RevModPhys.82.2313, PhysRevLett.93.130501, doi:10.1126/science.1231930}. The implementation of quantum gates often involves the use of quantum optimal control to find fields that can suitably manipulate a quantum system \cite{doi:10.1080/00018732.2010.505452, 10.1063/1.3124084, RevModPhys.76.1037}. In these cases, the goal is a unitary transformation corresponding to a quantum gate. Quantum optimal control has been studied both theoretically \cite{OC_the_1, OC_the_2, OC_the_3, OC_the_4, OC_the_5, OC_the_6, OC_the_7} and experimentally \cite{OC_exp_1, OC_exp_2, OC_exp_3, OC_exp_4, OC_exp_5, OC_exp_6} in many contexts. Particularly important to quantum computing implementations, quantum control has been extended to create robust controls that achieve high fidelities even in the presence of noise and errors \cite{kosut2022robust, PhysRevA.104.053118, daems2013robust, dridi2020optimal, ge2020robust}.   

Despite gate creation via optimal control being an active area of research, the mechanisms behind \textit{how} quantum gates are created are not fully understood. The pulses generated by optimal control algorithms can be complicated and there is often little physical insight behind the quantum dynamics that are occurring. A definition of quantum control mechanism was proposed by Mitra and Rabitz that decomposed the evolution of a quantum system into a set of quantum pathways \cite{abhra_1}. Each quantum pathway has a corresponding complex amplitude, and the collective significant amplitudes will constructively and/or destructively interfere to produce the dynamics of a quantum system. These pathways and pathway amplitudes provide a systemic and quantitative way of understanding the mechanism behind the dynamics that occur when the control pulse is active. Additionally, Mitra and Rabitz proposed an efficient way of calculating pathway amplitudes known as Hamiltonian encoding \cite{abhra_1} which has recently been further improved upon \cite{OHPE}. Pathway mechanism analysis has been applied to a variety of problems both in simulations \cite{abhra_2, abhra_3, sharp_1, PhysRevA.93.053407, mitra2004mechanistic} and experiments \cite{abhra_4, rey2010laboratory, rdc}, and this paper extends the prior work through simulations to reveal the mechanism behind the creation of quantum gates. The work described in this paper was performed with closed, noiseless model systems of a small number of qubits. This paper provides several illustrative examples of pathway mechanism analysis to demonstrate the physical insights that can be gleaned. A researcher may adopt the same mechanistic analysis tools for any desired gate created by a particular optimal control.      
    
The background behind mechanistic quantum pathways and Hamiltonian encoding is provided in Section \ref{sec:pathway_and_HE}. Section \ref{sec:mechanism_gates} {presents} the mechanism analysis of a single qubit X-gate, two CNOT gates (i.e.~each CNOT gate was created with a distinct optimal field), and a SWAP gate. The mechanisms behind other gates were examined as well and those depicted here were chosen as representative of the type of information that may be gained. Finally, concluding remarks and potential future directions are discussed in Section \ref{sec:conclusion}. 

\section{Determining Mechanistic Quantum Pathways through Hamiltonian Encoding}\label{sec:pathway_and_HE}

This section serves to review previous work as background on extracting mechanistic pathways via Hamiltonian encoding \cite{abhra_1}. A quantum pathway between two states $\ket{a}$ and $\ket{b}$ is a sequence of states $\ket{l_1},\dots,\ket{l_{n-1}}$ connecting the two: $a \tto l_1 \tto \dots \tto l_{n-1}\tto b$. The corresponding complex pathway amplitudes are obtained by interpreting terms in the Dyson series expansion of time-evolution operator. Hamiltonian encoding is a procedure to calculate pathway amplitudes by modulating the Hamiltonian and decoding the subsequent evolution instead of direct computation. More details behind quantum pathways and Hamiltonian encoding are in \cite{abhra_1, abhra_2, sharp_1}, and further details on computation and implementation can be found in \cite{OHPE}.

Quantum pathways connecting states $\ket{a}$ and $\ket{b}$ only provide mechanistic information on a particular matrix element $\mel{b}{U(T)}{a}$ of the time evolution operator at a final time $T$. A quantum gate is a unitary transform $U$ and one may be interested in how several or all of the matrix elements of $U$ were formed. In these cases, mechanism analysis permits considering each matrix element of $U$ separately.

\subsection{Pathways amplitudes extracted from the Dyson Series}\label{ssec:pathways}

The $n$-qubit quantum systems addressed in this paper will have Hamiltonians which can be written as $H(t) = H_0 + H_c(t)$ where $H_0$ is a constant drift term and $H_c(t)$ is a time-varying control term. For a basis, we will always use the $2^n$ elements of the computational basis $\ket i$ for $i = 0, \dots, 2^n-1$ where $i$ is typically expressed in base 2, corresponding each digit with a qubit.

To define a quantum pathway amplitude, we will switch to the interaction picture. Let $V(t) = \exp(iH_0t/\hbar) H_c(t) \exp(-iH_0t/\hbar)$ be defined as the interaction Hamiltonian. Then the Schr\"odinger equation for the time-evolution operator $U(t)$ is 
\begin{equation}
    i\hbar \dv{t}U(t) = V(t) U(t),\ U(0) = I. \label{eq:schro_eq}
\end{equation}
The formal solution for $U(T)$ is given by the Dyson series:
\begin{align*}\label{eq:dyson_full}
     U&(T) = I + \qty(\frac{-i}{\hbar})\int_{0}^{T} V(t_1) \dd{t_1}\\  \notag
     &+ \qty(\frac{-i}{\hbar})^2\int_{0}^{T}\! \int_{0}^{t_2} V(t_2)  V(t_1) \dd{t_1} \dd{t_2}\\ \notag
     &+ \qty(\frac{-i}{\hbar})^3\int_{0}^{T}\! \int_{0}^{t_3}\! \int_{0}^{t_2} V(t_3)  V(t_2)  V(t_1) \dd{t_1} \dd{t_2}\dd{t_3}\\ \notag
     &+\ \cdots. \nmberthis
\end{align*}
If we insert the identity $I = \sum_{i=0}^{2^n-1} \dyad{i}$ between every matrix product and adopt the notation $U_{ba} = \mel{b}{U(T)}{a}$ and $v_{ji}(t) = - \frac{i}{\hbar} \mel{j}{V(t)}{i}$, then Eq. \eqref{eq:dyson_full} can be expressed as 
\begin{equation} \label{eq:dyson_elements}
    U_{ba} = \sum_{n=0}^\infty \sum_{l_{n-1}=0}^{2^n-1}\cdots\sum_{l_1=0}^{2^n-1} U_{ba}^{n(l_1,\dots,l_{n-1})}
\end{equation}
where
\begin{align} \label{eq:pathway_amps}
        U^{n(l_1,\dots,l_{n-1})}_{ba}\hspace{30pt}& \nonumber \\ 
        \equiv \int_0^T\int_0^{t_n}\cdots\int_0^{t_2} &v_{bl_{n-1}}(t_n)v_{l_{n-1}l_{n-2}}(t_{n-1})\cdots \nonumber \\ 
        & v_{l_1a}(t_1) \dd{t_1}\cdots\dd{t_{n-1}}\dd{t_n}.
\end{align}
The index $n$ in Eqs. \eqref{eq:dyson_elements} and \eqref{eq:pathway_amps} denotes the \textit{order} of the term {in the Dyson expansion}, i.e. the number of transitions induced by the interaction $V${, and} the parenthesis $(l_1,\dots,l_{n-1})$ lists the $n-1$ intermediate states. An $n$-th order \textit{pathway} between two states $\ket{a}$ and $\ket{b}$ is a sequence of $n$ transitions $a \tto l_1 \tto\dots\tto l_{n-1}\tto b$ through $n-1$ intermediate states $\ket{l_i}$. Equation \eqref{eq:pathway_amps} gives the complex \textit{pathway amplitude} for a given pathway. Pathway amplitudes with greater magnitudes have larger contributions to the dynamics and the phases of the amplitudes allow for various pathways to constructively and/or destructively interfere. For a given control problem there normally may be infinitely many pathways, but the Dyson series always converges for a bounded Hamiltonian \cite{abhra_1} and the core mechanism is dominated by the largest amplitudes and the interference pattern between them. 

In the following analyses, it will be useful to group pathways with similar attributes into \textit{pathway classes}. The amplitude of a pathway class is the sum of all the individual pathway amplitudes within the class. The two cases used in this paper will be referred to as either Hermitian or non-Hermitian pathway classes. A \textit{non-Hermitian pathway class}, denoted with a $NH$ superscript, is a set of pathways that only differ by \textit{time-sequencing}: the order in which transitions occur in a pathway. For example, the pathways $00 \tto 01 \tto 00 \tto 10 \tto 00$ and $00 \tto 10 \tto 00 \tto 01 \tto 00$ belong to the same non-Hermitian pathway class $[00 \tto 01 \tto 00 \tto 10 \tto 00]^{NH}$. Many non-Hermitian pathway classes like $[00 \tto 01 \tto 11]^{NH}$ may only contain one pathway. A \textit{Hermitian pathway class}, denoted with an $H$ superscript, denotes the set of pathways that differ only by time-sequencing and \textit{backtracking.} Backtracking occurs when a pathway includes a transition from state $\ket i$ to $\ket j$ and later a transition from $\ket j$ to $\ket i$. Backtracking includes Rabi flopping: when a pathway involves a transition from $\ket i$ to $\ket j$ and immediately thereafter a transition from $\ket j$ back to $\ket i$. For example, both $00 \tto 10$ and $00 \tto 01 \tto 00 \tto 10$ are members of the $[00 \tto 10]^H$ Hermitian pathway class. Hermitian pathway classes do not provide as much detailed information about the mechanism as the non-Hermitian case, but they are important in often providing an adequate coarse-grained description of the mechanism.

\subsection{Hamiltonian Encoding} \label{ssec:Hamitlonian encoding}

In principle, each pathway amplitude $U_{ba}^{n(l_1,...,l_{n-1})}$ may be computed directly via Eq.~\eqref{eq:pathway_amps}, but this becomes infeasible as the order $n$ increases: the number of possible pathways rises exponentially due to the growing number of intermediate states $(l_1, \dots, l_{n-1})$. Instead of directly computing these integrals, Hamiltonian encoding modulates the Hamiltonian in an additional time-like variable $s \geq 0$ and decodes the evolution of the subsequent modulated system to efficiently extract pathway amplitudes\cite{abhra_1, OHPE}.

\begin{figure}
    \centering
    \includegraphics[width=0.8\linewidth]{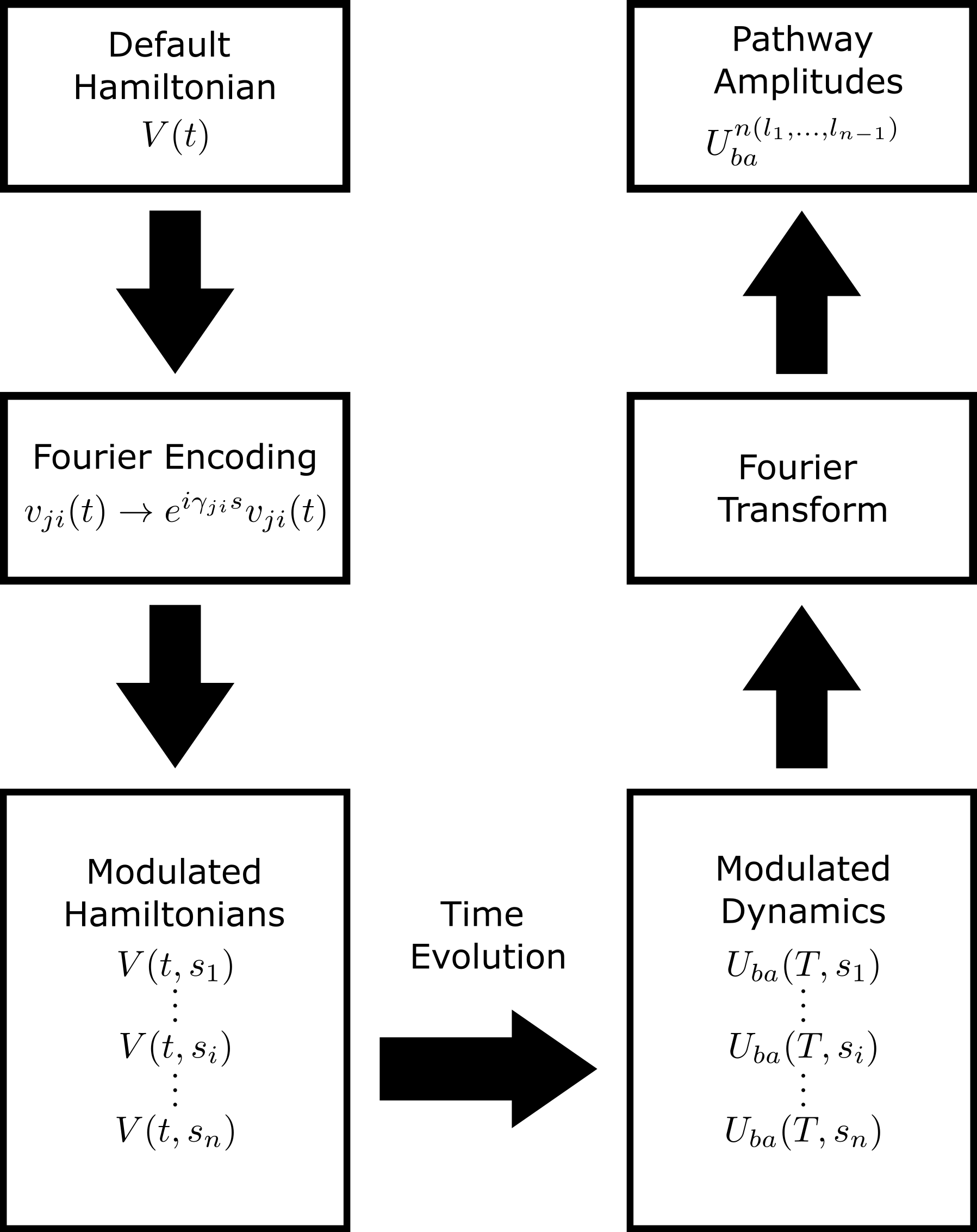}

    \caption{\justifying A diagram showing an outline of the Hamiltonian encoding process. The time-dependent Hamiltonian has its matrix elements modulated by complex exponentials in a time-like parameter $s$. At many values of $s$, the modulated Schr\"odinger equation is solved for the modulated system's evolution. The amplitudes of pathway classes are simultaneously extracted by decoding the modulated time evolution operators via a Fourier transform. }
    \label{fig:flowchart}
\end{figure}

Hamiltonian encoding is most commonly implemented as \textit{Fourier encoding} which entails modulating each matrix element of the Hamiltonian by multiplying it by a complex exponential
\begin{equation} \label{eq:fourier_enc}
    v_{ji}(t) \rightarrow v_{ji}(t;s) = e^{i \gamma_{ji} s} v_{ji}(t)
\end{equation}
where each element is assigned a frequency $\gamma_{ji}$, whose value bears no connection to any physical frequencies of the system. This encoding gives the modulated Schr\"odinger equation:
\begin{equation}\label{eq:modulated_schr}
    \dv{U(t; s)}{t} = \pmqty{e^{i\gamma_{11}s}v_{11}(t) & \cdots & e^{i\gamma_{1d}s}v_{1d}(t) \\ \vdots & \ddots & \vdots \\ e^{i\gamma_{d1}s}v_{d1}(t) & \cdots & e^{i\gamma_{dd}s}v_{dd}(t)}U(t; s).
\end{equation}
The modulated pathway amplitude $U^{n(l_1,\dots,l_{n-1})}_{ba}(s)$, at a final time $T$, for a pathway $a \tto l_1 \tto\dots\tto l_{n-1}\tto b$ can be written as 
\begin{align*} \label{eq:modulated_pathway_amp}
     &U^{n(l_1,\dots,l_{n-1})}_{ba}(s)\\ \notag
        &= \int_0^T\int_0^{t_n}\cdots\int_0^{t_2} v_{bl_{n-1}}(t_n)e^{i\gamma_{bl_{n-1}}s}\cdots\notag \\ 
        & \hphantom{\int_0^T\int_0^{t_n}\cdots\int_0^{t_2}}\hspace{15pt}v_{l_1a}(t_1)e^{i\gamma_{l_1a}s}\dd{t_1}\cdots\dd{t_{n-1}}\dd{t_n}  \\ 
        &= U^{n(l_1,\dots,l_{n-1})}_{ba}\, e^{i\gamma^{n(l_1,\dots,l_{n-1})}_{ba}s} \nmberthis
\end{align*}
where 
\begin{equation} \label{eq:gamma_pathway}
    \gamma^{{n(l_1,\dots,l_{n-1})}}_{ba} \equiv \gamma_{bl_{n-1}} + \dots + \gamma_{l_1a}.
\end{equation}
Each pathway has an associated frequency $\gamma^{{n(l_1,\dots,l_{n-1})}}_{ba}$. Akin to Eq.~\eqref{eq:dyson_elements}, the modulated transition amplitude between $\ket{a}$ and $\ket{b}$ at final time $T$ can be written as
\begin{equation} \label{eq:Abhra_linear_eq}
    U_{ba}(s) =\  \sum_\text{pathways}  U^{n(l_1,\dots,l_{n-1})}_{ba} \cdot e^{i\gamma^{{n(l_1,\dots,l_{n-1})}}_{ba}s} .
\end{equation}
This relation is a sum of complex sinusoidal terms in $s$ where each has an associated complex pathway amplitude. We know the possible pathway frequencies and desire to determine the transition amplitudes associated to them. Given sufficiently many sample points $s$ in Eq.~\eqref{eq:modulated_schr}, a Fourier transform of $U_{ba}(s)$ can be used to efficiently extract the pathway amplitudes $U^{n(l_1,\dots,l_{n-1})}_{ba}$ for a given frequency $\gamma^{{n(l_1,\dots,l_{n-1})}}_{ba}$. The main computational cost of Hamiltonian encoding is solving Eq.~\eqref{eq:modulated_schr} for different $s$-points. This procedure is laid out diagrammatically in Figure \ref{fig:flowchart}. 

The choice of frequencies in Eq.~\eqref{eq:fourier_enc} plays an important role in determining the nature of the mechanism information obtained from Hamiltonian encoding. Certain choices can yield amplitudes corresponding to either Hermitian or non-Hermitian pathways classes as defined in Section \ref{ssec:pathways}. \textit{Hermitian (H) encoding} is defined to be
\begin{equation} \label{eq:hermitian_encoding}
\begin{matrix}
        v_{ji}(t) \rightarrow v_{ji}(t) e^{i\gamma_{ji}s}\\
        \gamma_{ij} = - \gamma_{ji}.
\end{matrix}
\end{equation}
Using the constraint in Eq.~\eqref{eq:hermitian_encoding}, pathways that contain backtracking (i.e. a $i \tto j$ and a $j \tto i$ transition) have frequencies that cancel in Eq.~\eqref{eq:gamma_pathway}. All pathways that differ by only backtracking have the same pathway frequency $\gamma^{{n(l_1,\dots,l_{n-1})}}_{ba}$. Therefore, the associated amplitude for this frequency is exactly the Hermitian pathway class amplitude consisting of the sum of all constituent pathway amplitudes. Hermitian pathway classes carry the label (H) since the encoding in Eq.~\eqref{eq:hermitian_encoding} keeps the modulated Hamiltonian Hermitian.

Non-Hermitian pathway class amplitudes are calculated using \textit{non-Hermitian encoding}
which relaxes the second condition in Eq.~\eqref{eq:hermitian_encoding}:
\begin{equation} \label{eq:non_hermitian_encoding}
\begin{matrix}
        v_{ji} \rightarrow v_{ji} e^{i\gamma_{ji}s}\\
        \gamma_{ij} \neq -\gamma_{ji}.
\end{matrix}
\end{equation}
Since the associated frequencies for $i \tto j$ and $j \tto i$ do not cancel, pathways that differ by backtracking will have different frequencies and therefore have distinguishable pathway amplitudes. Since the sum in Eq.~\eqref{eq:gamma_pathway} is commutative, non-Hermitian encoding does not distinguish time-sequencing information. Hamiltonians encoded with Non-Hermitian (NH) encoding are no longer Hermitian for general $s\neq 0$. However, the extracted mechanism with NH encoding still pertains to the physical (Hermitian) system dynamics.

In both H and NH encoding, there is freedom regarding which values to assign to each individual frequency $\gamma_{ji}$. Typically this is done to maximize the amount of pathway classes with unique associated frequencies. Additionally, recent work \cite{OHPE} has shown that not all transitions need to be modulated to determine full mechanistic information. This gives more efficient encoding procedures to obtain H and NH pathway classes, both of which are used in the illustrations in this paper. The reader is directed to \cite{OHPE} for further details on picking frequencies and efficient encodings.

\section{Mechanisms and Quantum Pathways for Creating Qubit Gates} \label{sec:mechanism_gates}

In this section, mechanism analysis is applied to qubit systems to illustrate the important insights that can be gained by the techniques outlined in Section \ref{sec:pathway_and_HE}. The systems investigated are gates arising from either single or two-qubit systems. Instead of viewing gate creation as a black-box operation, mechanism analysis reveals information about fundamental aspects of the quantum dynamics resulting from a given optimal control pulse. The quantitative mechanism for a prepared gate is expressed in terms of the quantum pathways involved. Qubit systems are ideal for assessing control mechanisms in this fashion because the computational basis is a discrete set of states as compared to a continuous basis (e.g., coherent/squeezed states of light or wave functions of a molecule in coordinate space). Discussions regarding such further applications beyond the scope of this paper are given in Section \ref{sec:conclusion}.

For the simulations performed in Sections \ref{ssec:cnot_gate} and \ref{ssec:swap_gate}, Eq.~\eqref{eq:modulated_schr} is solved numerically by approximating the control Hamiltonians as piecewise constant in time:
\begin{equation}
    U(T;s) \approx \prod_{n=1}^{T/\Delta t} \exp \qty(- \frac{i}{\hbar} V(n \Delta t; s) \Delta t).
\end{equation}
When $V(n \Delta t; s)$ is Hermitian, the matrix exponential is efficiently calculated by diagonalizing $V$. When $V(n \Delta t; s)$ is not Hermitian (i.e. for NH encoding) the matrix exponential is computed using a squaring and scaling method for numerical stability and accuracy \cite{exp}. The choice of frequencies and encoding procedures were performed according to the conventions and procedures in \cite{OHPE}.

Except for the X-gate, all control pulses were generated using the GRAPE algorithm \cite{GRAPE}, which attempts to minimize the cost function $| \tr\phantom{}\{  U^\dagger_\textrm{target} U(T) \}|$.

\subsection{X-Gate} \label{ssec:x_gate}

\begin{figure}[!]
    \includegraphics[width=\linewidth]{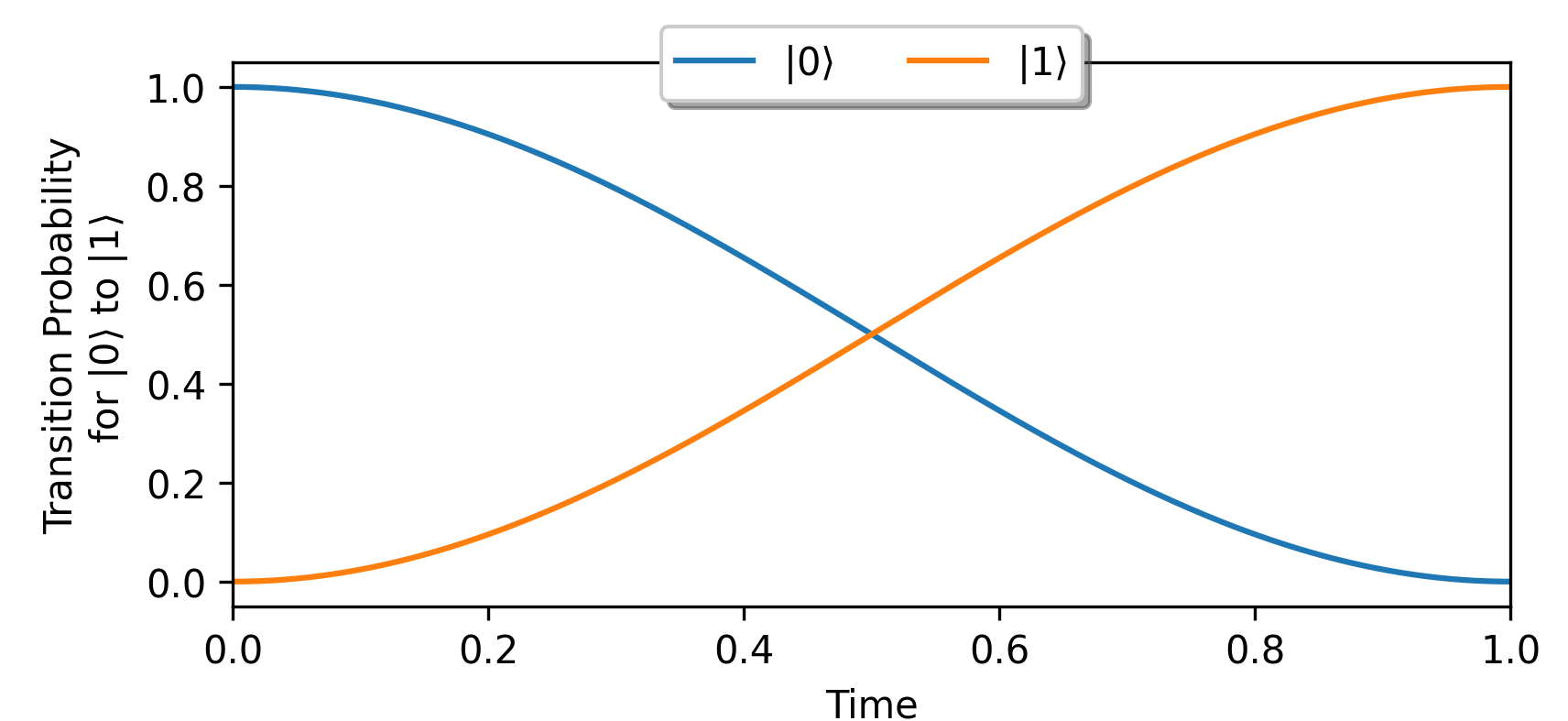}
    \caption{\justifying  Populations of the $\ket{0}$ and $\ket{1}$ states under the effect of a $\pi$-pulse for creating an $X$ gate with a Hamiltonian given by Eq.~\eqref{eq:Xgate-hamiltonian}. }
    \label{fig:Xgate_population}
\end{figure}

For the $X$ gate, we work in the rotating frame such that $H_0 = 0$. An X-gate is implemented as a $\pi$-pulse
\begin{equation} \label{eq:Xgate-hamiltonian}
    H_c(t) = \begin{cases}
        \pi S_x &  t < 1\\
        0 & t > 1
    \end{cases}
\end{equation}
where $S_x$ is the Pauli spin operator in the $x$ direction. The analysis will be performed assuming that the system starts in state $\ket{0}$, but the analysis starting in state $\ket{1}$ is analogous. Population plots depicting the evolution of this system are shown in Figure \ref{fig:Xgate_population}.

Given the simple monotone behavior of the population plots in Figure \ref{fig:Xgate_population}, it is tempting to suspect that the mechanism consists of the first-order and second-order perturbation theory terms, respectively, $U_{10}^{1}$ and $U_{00}^{2(1)}$. However, this speculation cannot be the case because with these two terms alone, the time evolution is not unitary. This simple example illustrates how population plots can hide the true mechanism, which inevitably involves higher-order terms.

This particular system is so simple that pathway amplitudes can be calculated analytically, thereby providing a good illustrative mechanism analysis without the need for Hamiltonian encoding. Given a state $\ket{0}$ or $\ket{1}$, the only possible full transition the system can make is to $\ket 1$ or $\ket{0}$, respectively. Therefore, all pathways connecting $\ket 0$ to $\ket{0}$ must have an even number of transitions with pathway amplitudes given by
\begin{align} \label{eq:Xgate_00_transition}
    U^n_{00} &= \begin{cases}
        0 & $n$\text{ odd}\\
        \displaystyle \frac{1}{n!} \qty(\frac{-i\pi}{2})^n & $n$\text{ even}\\
    \end{cases}
\end{align}
where $n$ is the order of the pathway. A similar result holds for pathways from $\ket0$ to $\ket 1$:
\begin{align} \label{eq:Xgate_10_transition}
    U^n_{10} &= \begin{cases}
        \displaystyle \frac{1}{n!} \qty(\frac{-i\pi}{2})^n & $n$\text{ odd}\\
        0 & $n$\text{ even}
    \end{cases}.
\end{align}
The pathways and their corresponding amplitudes are presented in Table \ref{tab:Xgate_pathways}.

\begin{table}[t]
\begin{center}
\subfloat{
\begin{tabular}{l c c} 
 \hline \hline
 Pathway & Magnitude & Phase \\ [0.5ex] 
 \hline
 \addlinespace[1pt]  $0$ & 1.000 &  0$^\circ$ \\
 \addlinespace[1pt]  $0\tto1\tto0$ & 1.234 & 180$^\circ$  \\
 \addlinespace[1pt]  $0\tto1\tto0\tto1\tto0$ & 0.254  &  0$^\circ$\\
 \addlinespace[1pt]  $0\tto1\tto0\tto1\tto0\tto1\tto0\phantom{\tto0~}$ &  0.021 & 180$^\circ$ \\ 
 $\cdots$ & $\cdots$ & $\cdots$ \\
 \hline
 Sum  & 0 & 0$^\circ$ \\ 
 \hline \hline
\end{tabular}}
\vspace{0.25cm}
\subfloat{
\begin{tabular}{l c c} 
 \hline \hline
 Pathway & Magnitude & Phase \\ [0.5ex] 
 \hline
 \addlinespace[1pt]  $0\tto1$ & 1.571 &  270$^\circ$ \\
 \addlinespace[1pt]  $0\tto1\tto0\tto1$ & 0.646 & 90$^\circ$  \\
 \addlinespace[1pt]  $0\tto1\tto0\tto1\tto0\tto1$ & 0.080  &  270$^\circ$\\
 \addlinespace[1pt]  $0\tto1\tto0\tto1\tto0\tto1\tto0\tto1$ &  0.005 & 90$^\circ$ \\ 
 $\cdots$ & $\cdots$ & $\cdots$ \\
 \hline
 Sum  & 1.000 & 270$^\circ$ \\ 
 \hline \hline
\end{tabular}}
\end{center}
\caption{\justifying  Pathways, amplitude magnitudes, and amplitude phases (in degrees here and later tables) of a $\pi$-pulse on a two-level X-Gate. The upper table includes pathways from $\ket 0$ to $\ket 0$ which perfectly destructively interfere since $U_{00}(T)=0$. The lower table includes pathways from $\ket 0$ to $\ket 1$ which 
both constructively and destructively interfere in a manner that results in the desired matrix element $|U_{10}(T)| = 1$ with an overall $270^\circ$ phase that is physically irrelevant.}
\label{tab:Xgate_pathways}
\end{table}

We return again to the prior remark about the time dependence in Figure~\ref{fig:Xgate_population} showing monotonically decreasing population in state $\ket 0$ and monotonically increasing population in state $\ket{1}$. At first glance, this might suggest that the dominant pathway is only $0 \tto 1$, but Eq.~\eqref{eq:Xgate_10_transition}  along with Table \ref{tab:Xgate_pathways} show the significance of numerous higher-order pathways corresponding to oscillations back and forth between the $\ket 0$ and $\ket 1$ states. At final time $T=1$, the $\ket{0}$ to $\ket{0}$ pathways fully destructively interfere and the $\ket{0}$ to $\ket{1}$ pathways both constructively and destructively interfere in a manner that leaves the whole system in the $\ket 1$ state. Even in the seemingly simple dynamics of this 2-level system, there is a clear dependence on higher-order interactions and interference patterns between the pathway amplitudes. This mechanistic information is inaccessible from the population plots in Figure \ref{fig:Xgate_population} alone and the situation is exasperated for higher dimensional and more complex gates, as will be demonstrated in the following subsections.

\subsection{CNOT Gate} \label{ssec:cnot_gate}

In this section, two different implementations of the CNOT gate 
\begin{equation} \label{eq: CNOT_definition}
    \text{CNOT} = \mqty(1 & 0 & 0 & 0\\
    0 & 1 & 0 & 0\\
    0 & 0 & 0 & 1\\
    0 & 0 & 1 & 0)
\end{equation}
are investigated to compare and contrast the distinct controls and their respective mechanisms that achieve the same goal. Control landscape analysis \cite{PhysRevA.79.013422} has shown the existence of multiple control solutions but the associated mechanisms have not been explored. Here we will choose two particular optimal controls for illustrative purposes. In the CNOT gate, the constant drift term of the full Hamiltonian is described by
\begin{equation}\label{eq:2q_h0}
    H_0 = \omega_1 (S_z \otimes \mathbb{I}) +\omega_2 (  \mathbb{I} \otimes S_z) +  J (S_z \otimes S_z) 
\end{equation}
where $S_z$ is the Pauli spin operator in the $z$ direction, $\mathbb I$ is the $2 \times 2$ identity matrix, $\omega_1$ and $\omega_2$ are the resonance offset frequencies, and $J$ is the J-coupling.  The control term takes the form of
\begin{equation} \label{eq:2q_hc}
    H_c(t)= \epsilon{_x}(t) (S_x \otimes \mathbb{I} + \mathbb{I} \otimes S_x ) + \epsilon{_y}(t) (S_y \otimes \mathbb{I} +  \mathbb{I} \otimes S_y) 
\end{equation}
where $S_x$ and $S_y$ are the Pauli spin operators in the $x$ and $y$ directions and $\epsilon{_x}(t)$ and $\epsilon{_y}(t)$ are control fields applied in the respective directions. This is a particular model for two coupled qubits, but other models could be similarly explored. The eigenstates of $H_0$ are the four two-qubit states: $\ket{00}$, $\ket{01}$, $\ket{10}$, and $\ket{11}$. In our convention, the first digit is the control qubit and the second digit is the target qubit\footnote{This terminology will also apply to intermediate time steps when the control Hamiltonian is non-zero.}. The subsequent mechanism analysis will only consider three matrix elements\footnote{In theory, up to 16 separate matrix elements could have been studied, but this extensive analysis is not needed to understand the control mechanism in this case.}: $\mel{00}{U(T)}{00}$ and $\mel{11}{U(T)}{10}$ respectively corresponding to not flipping the target qubit and flipping the target qubit as well as an additional zero matrix element $\mel{10}{U(T)}{10}$. The control Hamiltonian only permits single qubit flips at a time, i.e. $00\tto10$ and $11\tto 10$ are allowed but a direct transition like $00 \tto 11$ is not. The two CNOT gates considered here only differ in their optimal control fields $\epsilon{_x}(t)$ and $\epsilon{_x}(t)$ which were obtained by a gradient ascent algorithm respectively starting from distinct initial trial fields. For clarity, the two implementations of a CNOT gate we consider will be referred to as CNOT(i) and CNOT(ii) henceforth. The optimal control code added a physically irrelevant global phase of $135^\circ$ to both CNOT gates.

\subsubsection{CNOT(i) Gate}

\begin{figure}
    \includegraphics[width= \linewidth]{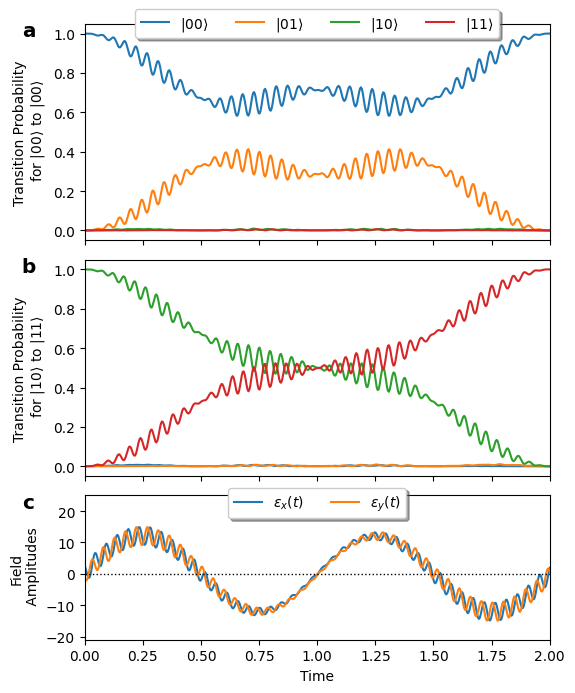}
    \caption{\justifying  (a) and (b) show the populations of all four computational basis states driven by the Hamiltonian in Eqs.~\eqref{eq:2q_h0} and \eqref{eq:2q_hc} under the effect of CNOT(i) acting on the initial states $\ket{00}$ and $\ket{10}$, respectively. (c) shows the two control fields $\epsilon{_x}(t)$ and $\epsilon{_y}(t)$ that implement CNOT(i). There is a clear qualitative relation between the high-frequency backtracking features in (a) and (b) and the oscillations in the two fields in (c).} 
    \label{fig:CNOT1}
\end{figure}

The first CNOT gate that will be investigated has population plots shown in Figures \ref{fig:CNOT1}a and \ref{fig:CNOT1}b for initial states $\ket{00}$ and $\ket{10}$ respectively. Although the basis consists of four states, only two population plots with distinct initial states are depicted with the key difference being that \ref{fig:CNOT1}a is concerned with mechanism analysis for $\mel{00}{U(T)}{00}$ and \ref{fig:CNOT1}b for $\mel{11}{U(T)}{10}$. The following mechanism results are similar to the matrix elements $\mel{01}{U(T)}{01}$ and $\mel{10}{U(T)}{11}$, so these will not be shown. All other matrix elements in the CNOT gate are zero and will exhibit complete destructive interference in the mechanism, which will be discussed later.

Before moving to the pathway mechanism analysis, we consider information that can be gleaned without the use of Hamiltonian encoding. The population plots in Figures \ref{fig:CNOT1}a and \ref{fig:CNOT1}b mainly show two states being present throughout the evolution in each case. When starting in $\ket{00}$ in Figure \ref{fig:CNOT1}a, the control fields initially bring the system into $\ket{01}$ before undoing this process in the second half of the pulse. In the second case of Figure \ref{fig:CNOT1}b starting in $\ket{10}$, the control fields increase the population in the target $\ket{11}$ population throughout the whole duration of the pulse, thereby finally leaving the system entirely in the $\ket{11}$ state. The high-frequency features in the populations are indications of Rabi-flopping, but no quantitative information can be gleaned about this matter without Hamiltonian encoding.

\begin{figure}
    \includegraphics[width=\linewidth]{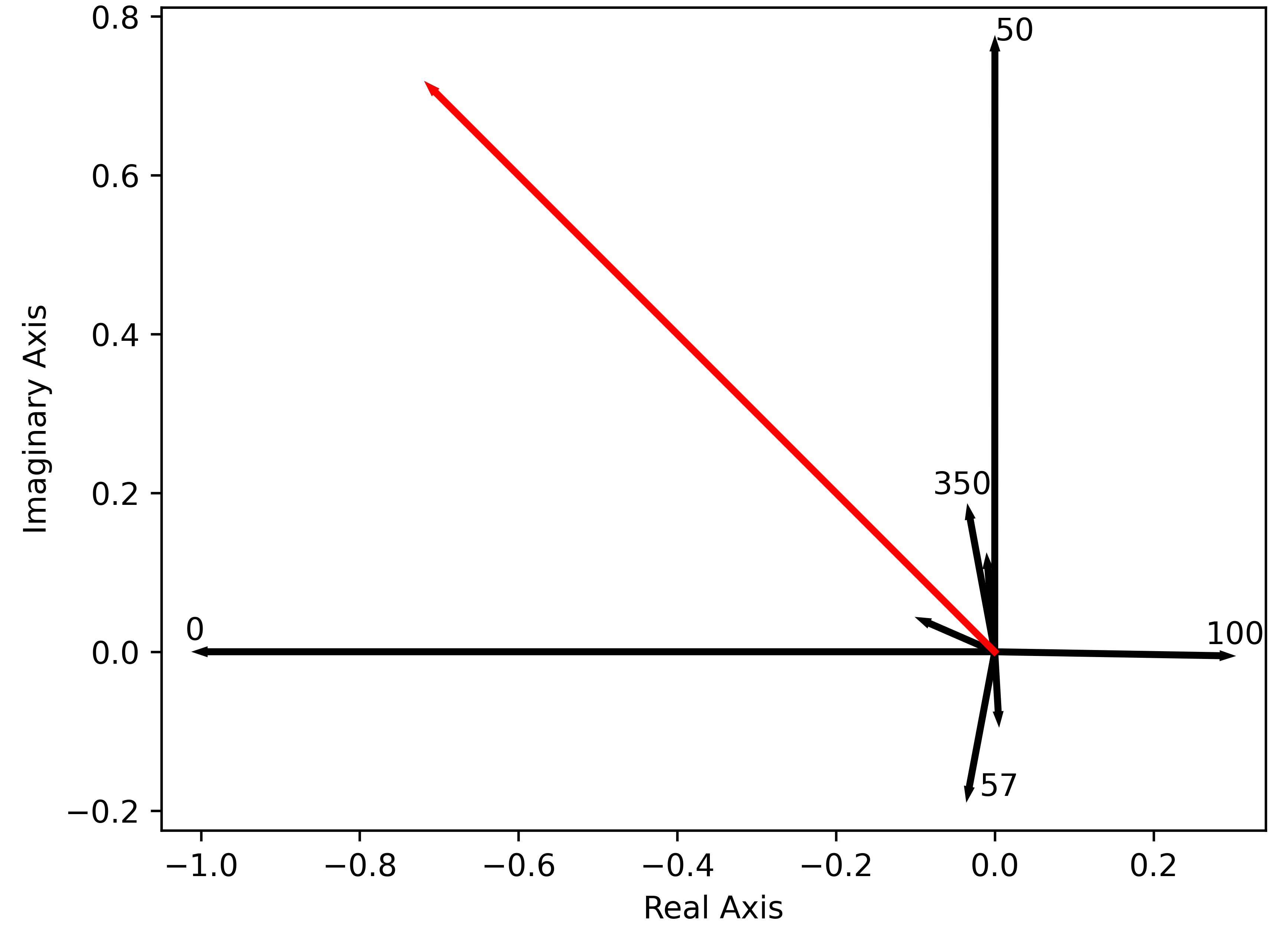}
    \caption{\justifying The NH pathway class amplitudes of CNOT(i) for pathways from $\ket{00}$ to $\ket{00}$. Each amplitude is drawn as a vector in the complex plane and pathway classes with large enough magnitudes are labeled with their associated frequency. The red vector is the sum of all pathway class amplitudes and has a magnitude equal to 1. These NH pathway classes are expanded upon in Table \ref{tab:CNOT1_pathways_00_00}.}
    \label{fig:CNOT1gate_arrow_plots_00_00}
\end{figure}

\begin{figure}
    \includegraphics[width=\linewidth]{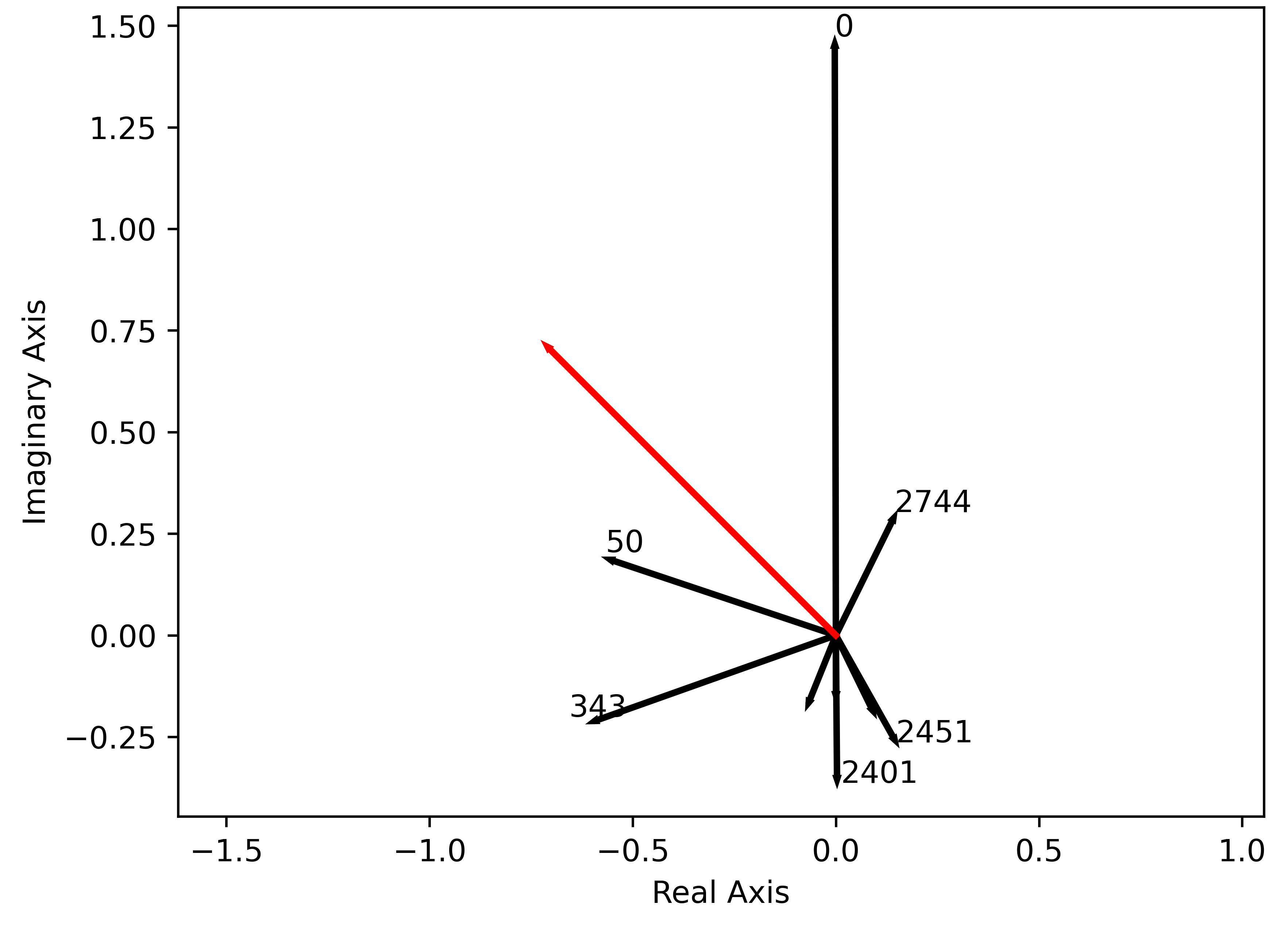}
    \caption{\justifying The NH pathway class amplitudes of CNOT(i) for pathways from $\ket{10}$ to $\ket{11}$. Each amplitude is drawn as a vector in the complex plane and pathway classes with large enough magnitudes are labeled with their associated frequency. The red vector is the sum of all pathway class amplitudes and has a magnitude equal to 1. These NH pathway classes are expanded upon in Table \ref{tab:CNOT1_pathways_10_11}.}
    \label{fig:CNOT1gate_arrow_plots_10_11}
\end{figure}

Non-Hermitian Hamiltonian encoding was performed on the system giving a set of associated pathway classes and corresponding amplitudes at the final time $T=2$. This four-dimensional system required only $~16000$ solves of Eq.~\eqref{eq:modulated_schr} for each initial state. Tables \ref{tab:CNOT1_pathways_00_00} and \ref{tab:CNOT1_pathways_10_11} give the dominant pathways that evolved the system from $\ket{00}$ to $\ket{00}$ and $\ket{10}$ to $\ket{11}$, respectively corresponding to Figures \ref{fig:CNOT1}a and \ref{fig:CNOT1}b. As each pathway class corresponds to a complex pathway amplitude, these amplitudes can be plotted as vectors in the complex plane for easy visualization as shown in Figures \ref{fig:CNOT1gate_arrow_plots_00_00} and \ref{fig:CNOT1gate_arrow_plots_10_11}. In Table \ref{tab:CNOT1_pathways_00_00} and subsequent tables, the pathway class amplitudes are ordered by their decreasing magnitude and labeled by their associated frequency $\gamma_{ba}^{n(l_1, \dots, l_{n-1})}$ which is conveniently also used to label the pathways class amplitudes in the complex plan plots (e.g., Figure \ref{fig:CNOT1gate_arrow_plots_00_00}).



\begin{table*}
    
\centering
\begin{tabular}{c l c c} 
 \hline
 \hline
 \addlinespace[2pt]$\gamma^{n(l_1,\dots,l_{n-1})}_{ba}$ & Non-Hermitian Pathway Class & Magnitude & Phase \\ [0.5ex] 
 \hline
 \addlinespace[1pt] 0 & $[00 ]^\text{NH}$ & 1.000 & 180$^\circ$  \\
\addlinespace[1pt] 50 & $[00\tto10\tto00]^\text{NH}$ & 0.764 & 90$^\circ$  \\
\addlinespace[1pt] 100 & $[00\tto10\tto00\tto10\tto00]^\text{NH}$ & 0.292 & -1$^\circ$  \\
\addlinespace[1pt] 57 & $[00\tto10\tto00\tto01\tto00]^\text{NH}$ & 0.181 & -101$^\circ$  \\
\addlinespace[1pt] 350 & $[00\tto01\tto11\tto01\tto00]^\text{NH}$ & 0.178 & 101$^\circ$   \\
\addlinespace[1pt] 14 & $[00\tto01\tto00\tto01\tto00]^\text{NH}$ & 0.113 & 95$^\circ$   \\
\addlinespace[1pt] 107 & $[00\tto10\tto00\tto10\tto00\tto01\tto00]^\text{NH}$ & 0.098 & 157$^\circ$  \\
\addlinespace[1pt] 7 & $[00\tto01\tto00]^\text{NH}$ & 0.084 & -87$^\circ$  \\
 $\cdots$ & & $\cdots$ & $\cdots$ \\
 \hline
 Sum &  & 1.000 & 135$^\circ$ \\ 
 \hline
 \hline
\end{tabular} 
\caption{\justifying Non-Hermitian pathway classes, amplitude magnitudes, and amplitude phases for creating CNOT(i) considering the gate matrix element $\mel{00}{U(T)}{00}$. All pathway classes with a magnitude greater than $0.083$ are listed.}\label{tab:CNOT1_pathways_00_00}
\begin{tabular}{c l c c} 
 \hline
 \hline
 \addlinespace[2pt]$\gamma^{n(l_1,\dots,l_{n-1})}_{ba}$ & Non-Hermitian Pathway Class & Magnitude & Phase \\ [0.5ex] 
 \hline
 \addlinespace[1pt] 0   & $[10 \tto 11]^\text{NH}$ & 1.455 & 90$^\circ$  \\
\addlinespace[1pt] 343  & $[10 \tto 11 \tto 01 \tto 11]^\text{NH}$ & 0.631 & -160$^\circ$  \\
\addlinespace[1pt] 50   & $[10 \tto 00 \tto 10 \tto 11]^\text{NH}$ & 0.586 &  162$^\circ$  \\
\addlinespace[1pt] 2401 & $[10 \tto 11 \tto 10 \tto 11]^\text{NH}$ & 0.355 & -90$^\circ$ \\
\addlinespace[1pt] 2744 & $[10 \tto 11 \tto 10 \tto 11 \tto 01 \tto 11]^\text{NH}$ & 0.321 & 64$^\circ$   \\
\addlinespace[1pt] 2451 & $[10 \tto 00 \tto 10 \tto 11 \tto 10 \tto 11]^\text{NH}$ & 0.296 & -61$^\circ$   \\
\addlinespace[1pt] 686  & $[10 \tto 11 \tto 01 \tto 11 \tto 01 \tto 11]^\text{NH}$ & 0.207 & -64$^\circ$  \\
\addlinespace[1pt] 100  & $[10 \tto 00 \tto 10 \tto 00 \tto 10 \tto 11]^\text{NH}$ & 0.180 &-112$^\circ$ \\
\addlinespace[1pt] 4802 & $[10 \tto 11 \tto 10 \tto 11 \tto 10 \tto 11]^\text{NH}$ & 0.149 & -90$^\circ$   \\
 $\cdots$ & & $\cdots$ & $\cdots$ \\
 \hline
 Sum &  & 1.000 & 135$^\circ$ \\ 
 \hline
 \hline
\end{tabular}
\caption{\justifying Non-Hermitian pathway classes, amplitude magnitudes, and amplitude phases for creating CNOT(i) considering the gate matrix element $\mel{11}{U(T)}{10}$. All pathway classes with a magnitude greater than $0.13$ are listed. }
\label{tab:CNOT1_pathways_10_11}
\end{table*}

Since NH pathway classes reveal backtracking information, they are ideal for providing quantitative information regarding the oscillations found in the population plots. Indeed, upon examining Tables \ref{tab:CNOT1_pathways_00_00} and \ref{tab:CNOT1_pathways_10_11}, if backtracking was ignored, the pathways would all simplify to the $[00\tto 00]^H$ or $[10 \tto 11]^H$ Hermitian pathway classes, respectively. The large amount of backtracking is not surprising given the sinusoidal bumps in the population plots. Even the simplest systems can exhibit backtracking as was seen in the mechanism of the X-gate example in Section \ref{ssec:x_gate} where the backtracking was not evident in the population plots of Figure \ref{fig:Xgate_population}. In the CNOT(i) case, the states that the pathways used during backtracking are surprising. Consider the $10\tto11$ pathways in Table \ref{tab:CNOT1_pathways_10_11}, there are many (specifically the 2nd and 3rd most significant by magnitude) with $\ket{01}$ and $\ket{00}$  as intermediate states. However, this is not what one would expect from the population plot in Figure \ref{fig:CNOT1}b where the $\ket{00}$ and $\ket{01}$ states have a low presence throughout the entirety of the time duration. This type of phenomenon is not unique to this particular CNOT(i) gate. For example, it is widely encountered in stimulated Raman adiabatic passage (STIRAP) \cite{STIRAP}. STIRAP, in its simplest form, addresses a state-to-state population transfer problem with the use of a third intermediate state. The third state is often not visible in population plots but is fundamental in the dynamics, similar to the $\ket{00}$ and $\ket{01}$ states of CNOT(i) being investigated here.

The reason why states can show up in pathways but not appear in the final population is due to a high degree of destructive interference occurring in the mechanism. Returning to the $10\tto11$ example, consider the pathways with frequencies 343, 2744, and 686. All of these pathways as shown in Table \ref{tab:CNOT1_pathways_10_11} have $\ket{01}$ as an intermediate state. Identifying these pathways in the vector plot in Figure \ref{fig:CNOT1gate_arrow_plots_10_11} (the 686 pathway is unlabeled but it points in essentially the same direction as the 2451 pathway) reveals that they are all misaligned. Therefore they destructively interfere, leaving essentially no $\ket{01}$ population in the plot of Figure \ref{fig:CNOT1}b. When combined with all the other pathway classes that do not use the $\ket{01}$ state, the sum of all the pathway amplitudes still gives a value with a magnitude near 1. Hence at the final time, the system will be entirely in the $\ket{11}$ state. Similar statements are true for the use of $\ket{00}$ as an intermediate state as well. In the case of CNOT(i), the destructive interference of subsets of pathways allows for states to be important to the mechanism despite not physically showing up in the system population plots to any significant degree. 

\begin{figure}
    \includegraphics[width=\linewidth]{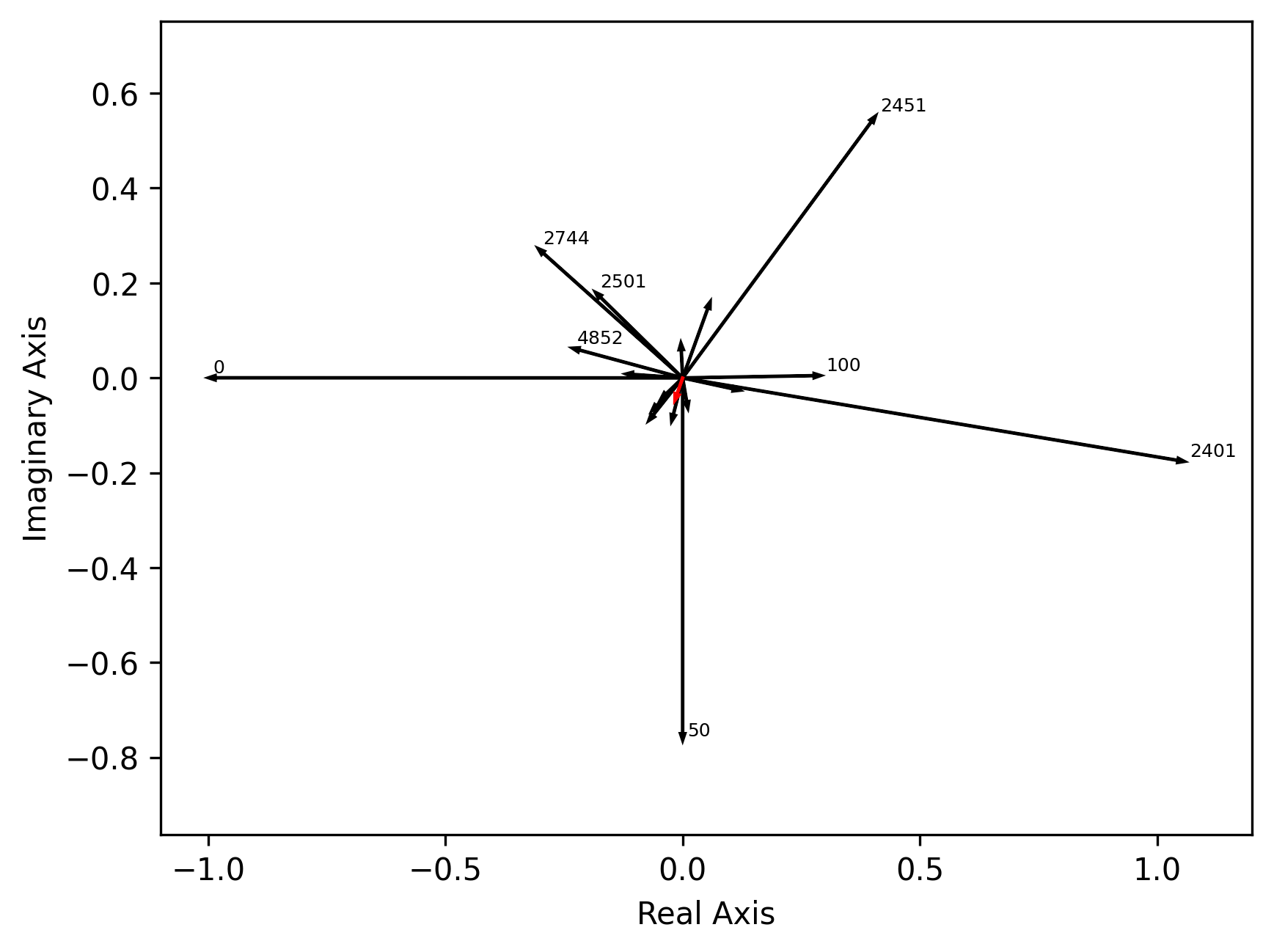}
    \caption{\justifying The NH pathway class amplitudes of CNOT(i) for pathways from $\ket{10}$ to $\ket{10}$. Each amplitude is drawn as a vector in the complex plane and pathway classes with large enough magnitudes are labeled with their associated frequency. The red vector is the sum of all pathway class amplitudes and has a magnitude nearly equal to 0. These NH pathway classes are expanded upon in Table \ref{tab:CNOT1_pathways_10_10}.}
    \label{fig:CNOT1gate_arrow_plots_10_10}
\end{figure}

In our analysis of CNOT(i) above, we examined two non-zero matrix elements, $\mel{00}{U(T)}{00}$ and $\mel{11}{U(T)}{10}$, of the unitary operator. Similar to the X-gate considered in Section \ref{ssec:x_gate}, zero matrix elements are equally important to consider for the mechanism analysis of the CNOT(i) gate. We consider the $\mel{10}{U(T)}{10}$ matrix element. Table \ref{tab:CNOT1_pathways_10_10} shows the pathways and Figure \ref{fig:CNOT1gate_arrow_plots_10_10} gives the associated vector plot. The mechanism of this matrix element, which should be zero from Eq.~\eqref{eq: CNOT_definition}, has similar features to the non-zero matrix elements discussed (such as pathways involving states with little presence in the population plots), but with the core difference that total sum of pathway amplitudes is near zero at the final time $T$ instead of a number with norm 1. As an example of the similarity, the pathways with frequencies 50, 2451, and 100 for the $\mel{10}{U(T)}{10}$ matrix element use $\ket{00}$ and $\ket{01}$ as intermediate states similar to what we saw before for the $\mel{11}{U(T)}{10}$ matrix element.

\begin{table*}
    
\centering
\begin{tabular}{c l c c} 
 \hline
 \hline
 \addlinespace[2pt]$\gamma^{n(l_1,\dots,l_{n-1})}_{ba}$ & Non-Hermitian Pathway Class & Magnitude & Phase \\ [0.5ex] 
 \hline
\addlinespace[1pt] 2401 & $[10\tto11\tto10]^\text{NH}$                          & 1.073 & -9$^\circ$ \\
\addlinespace[1pt] 0    & $[10 ]^\text{NH}$                                     & 1.000 & 180$^\circ$ \\
\addlinespace[1pt] 50   & $[10\tto00\tto10]^\text{NH}$                          & 0.764 & -90$^\circ$ \\
\addlinespace[1pt] 2451 & $[10\tto00\tto10\tto11\tto10]^\text{NH}$              & 0.686 & 54$^\circ$ \\
\addlinespace[1pt] 2744 & $[10\tto11\tto01\tto11\tto10]^\text{NH}$              & 0.410 & 138$^\circ$ \\
\addlinespace[1pt] 100  & $[10\tto00\tto10\tto00\tto10]^\text{NH}$              & 0.292 & 1$^\circ$ \\
\addlinespace[1pt] 2501 & $[10\tto00\tto10\tto00\tto10\tto11\tto10]^\text{NH}$  & 0.259 & 136$^\circ$ \\
\addlinespace[1pt] 4852 & $[10\tto00\tto10\tto11\tto10\tto11\tto10]^\text{NH}$  & 0.242 & 165$^\circ$ \\
 $\cdots$ & & $\cdots$ & $\cdots$ \\
 \hline
 Sum &  & 0.002 & 71$^\circ$ \\ 
 \hline
 \hline
\end{tabular} 
\caption{\justifying Non-Hermitian pathway classes, amplitude magnitudes, and amplitude phases for creating CNOT(i) considering the gate matrix element $\mel{10}{U(T)}{10}$ for $T=2$. All pathway classes with a magnitude greater than $0.2$ are listed. As expected from Eq.~\eqref{eq: CNOT_definition}, the matrix element is essentially zero and this was achieved through total destructive interference of the collective pathway amplitudes.}
\label{tab:CNOT1_pathways_10_10}

\end{table*}

\subsubsection{Second CNOT(ii) Gate}

\begin{figure}[t]
    \includegraphics[width=\linewidth]{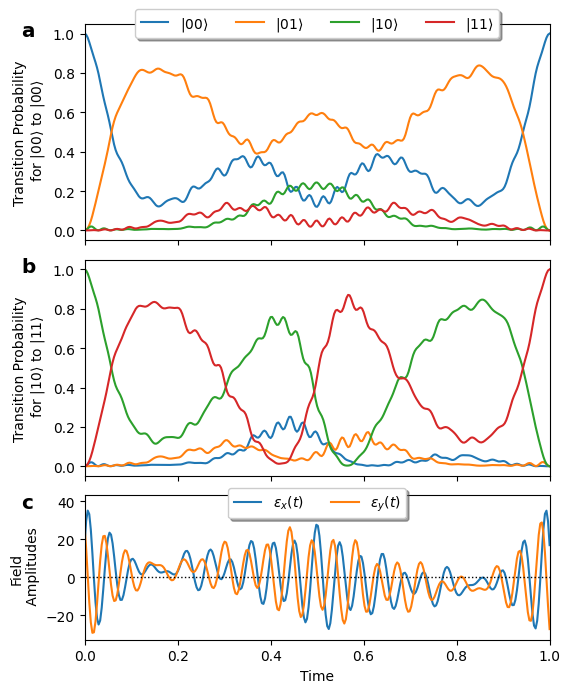}
    \caption{\justifying { (a) and (b) show the populations of all four computational basis states driven by the Hamiltonian in Eqs.~\eqref{eq:2q_h0} and \eqref{eq:2q_hc} under the effect of CNOT(ii) acting on the initial states either $\ket{00}$ or $\ket{10}$, respectively. (c) shows the two control fields $\epsilon{_x}(t)$ and $\epsilon{_y}(t)$ that implement CNOT(ii)}}
    \label{fig:CNOT2}
\end{figure}

The second CNOT(ii) gate investigated utilizes the same Hamiltonian system as the first gate with the only difference being a distinct set of optimal control fields $\epsilon_x(t)$ and $\epsilon_y(t)$. Additionally, for CNOT(ii) a final time of $T=1$ is used, consistent with the controls having higher fluence than CNOT(i). Population plots of the system starting in states $\ket{00}$ and $\ket{10}$ are depicted in Figures \ref{fig:CNOT2}a and \ref{fig:CNOT2}b respectively. The latter population plots, compared to the analogous plots in Figures \ref{fig:CNOT1}a and \ref{fig:CNOT1}b, indicate more dynamical complexity. Both cases for CNOT(ii) starting in either $\ket{00}$ or $\ket{10}$ have significant populations in all four states throughout most of the time interval whereas in the previous cases of CNOT(i), two states stayed mostly dormant (i.e.~as viewed through their population plots alone) during the entire pulse. However in a similar manner to CNOT(i), for the $\ket{00}$ initial state, the second half of the population evolution essentially reverses the evolution of the first half, while for the $\ket{10}$ initial state, the population evolution is more complex consistent with the initial and final states being distinct. 

It is unclear from the dramatically different plots in Figures \ref{fig:CNOT1}a and \ref{fig:CNOT1}b with respect to those of \ref{fig:CNOT2}a and \ref{fig:CNOT2}b whether the underlying mechanisms of CNOT(i) and CNOT(ii) are distinct. For example, although new states show up in the population plots in Figure \ref{fig:CNOT2}, it is possible that these states only appear in pathways with backtracking while not destructively canceling out as in the last analysis of CNOT(i). To answer, this question H encoding was performed on both systems to give Hermitian pathway classes and their corresponding amplitudes. Although NH encoding gives a fuller assessment of mechanistic detail, the Hermitian pathway classes ignore any backtracking information and are therefore ideal for determining if the CNOT(i) and CNOT(ii) gate mechanisms are fundamentally different. The results of the Hamiltonian encoding for both gates are given in Tables \ref{tab:CNOT2_pathways_00_00} and \ref{tab:CNOT2_pathways_10_11}.

\begin{table*}
\centering
\begin{tabular}{l c c c c} 
 \hline
 \hline
 & \multicolumn{2}{c}{{CNOT(i)}} & \multicolumn{2}{c}{{CNOT(ii)}}\\
 \addlinespace[1pt] Hermitian Pathway Class \hspace{7pt} & Magnitude & Phase \hspace{7pt} & Magnitude & Phase \\ [0.5ex] 
 \hline
 \addlinespace[1pt]  $[00\tto00]^\text{H}$ & 0.9997 & 136$^\circ$ & 0.934 & 154$^\circ$  \\
 \addlinespace[1pt]  $[00\tto01\tto11\tto10\tto00]^\text{H}$ & 0.0093 & 47$^\circ$ & 0.163 & 70$^\circ$  \\
 \addlinespace[1pt]  $[00\tto10\tto11\tto01\tto00]^\text{H}$ & 0.0093 & 47$^\circ$ & 0.162 & 66$^\circ$  \\
 $\cdots$ & $\cdots$ & $\cdots$ & $\cdots$ & $\cdots$ \\
 \hline
 Sum  & 1.000 & 135$^\circ$  & 1.000 & 135$^\circ$ \\ 
 \hline
 \hline
\end{tabular}
\caption{\justifying Hermitian pathway classes, amplitude magnitudes, and amplitude phases of both CNOT gates (i) and (ii) to create the matrix element $\mel{00}{U(T)}{00}$. All pathway classes with a magnitude greater than $0.001$ are listed.}
\label{tab:CNOT2_pathways_00_00}
\end{table*}

\begin{table*}
\centering
\begin{tabular}{l c c c c} 
 \hline
 \hline
 & \multicolumn{2}{c}{{CNOT(i)}} & \multicolumn{2}{c}{{CNOT(ii)}}\\
 \addlinespace[1pt] Hermitian Pathway Class \hspace{7pt} & Magnitude & Phase \hspace{7pt} & Magnitude & Phase \\ [0.5ex] 
 \hline
 \addlinespace[1pt]  $[10\tto11]^\text{H}$ & 0.9997 & 134$^\circ$ & 0.951 & 121$^\circ$  \\
 \addlinespace[1pt]  $[10\tto00\tto01\tto11]^\text{H}$ & 0.0059 & -171$^\circ$ & 0.129 & -152$^\circ$  \\
 \addlinespace[1pt]  $[10\tto11\tto01\tto00\tto10\tto11]^\text{H}$ & 0.0058 & -103$^\circ$ & 0.118 & -157$^\circ$  \\
 $\cdots$ & $\cdots$ & $\cdots$ & $\cdots$ & $\cdots$ \\
 \hline
 Sum  & 1.000 & 135$^\circ$  & 1.000 & 135$^\circ$ \\ 
 \hline
 \hline
\end{tabular}
\caption{\justifying Hermitian pathway classes, amplitude magnitudes, and amplitude phases of both CNOT gates to create the matrix element $\mel{11}{U(T)}{10}$. All pathway classes with a magnitude greater than $0.001$ are listed.}
\label{tab:CNOT2_pathways_10_11}
\end{table*}

As can be seen in Tables \ref{tab:CNOT2_pathways_00_00} and \ref{tab:CNOT2_pathways_10_11}, in either starting state with both CNOT(i) and CNOT(ii), there is one H pathway class with a significantly larger amplitude than the others. In the NH analysis performed in the previous section all the pathways depicted in Tables \ref{tab:CNOT1_pathways_00_00} and \ref{tab:CNOT1_pathways_10_11} were members of the $[00\tto 00]^\text{H}$ and $[10 \tto 11]^\text{H}$ pathway classes, respectively. In this regard, both CNOT gates at first glance have a fairly similar mechanism at the H pathway class level, with distinctions appearing only when backtracking is considered. 

Important differences in the mechanism between CNOT(i) and CNOT(ii) arise when investigating the non-dominate H pathway classes. For the $\mel{00}{U(T)}{00}$ case in Table \ref{tab:CNOT2_pathways_00_00}, the pathways classes involved are $[00\tto01\tto11\tto10\tto00]^\text{H}$ and $[00\tto10\tto11\tto01\tto00]^\text{H}$. In the CNOT(i), these pathways had a very small role. The $[00 \tto 00]^\text{H}$ pathway class already had 99.97\% of the magnitude for the total sum. However, in the CNOT(ii), the magnitudes of $[00\tto01\tto11\tto10\tto00]^\text{H}$ and $[00\tto10\tto11\tto01\tto00]^\text{H}$ are more than 15 times larger then for CNOT(i). A similar scenario occurs for the pathway classes in Table \ref{tab:CNOT2_pathways_10_11} to create $\mel{11}{U(T)}{10}$. The significance of these more complicated pathway classes implies that the mechanism CNOT(ii) is fundamentally different from CNOT(i). CNOT(ii) is not simply CNOT(i) but with different backtracking, instead, it utilizes fundamentally new routes and pathways to reach the target states.

These findings can be extended to analyze the mechanisms of other gates. A main point shown above is that different control fields creating the same final gate can have different mechanisms. Additionally, while two control pulses might look different, they could possibly lead to very similar mechanisms and system properties. These situations of similar or different mechanisms will be important to assess when desired ancillary properties are considered such as time-optimal control or robustness to noise. A clear general distinction is evident in the control fields of CNOT(i) in Figure~\ref{fig:CNOT1}c and CNOT(ii) in Figure~\ref{fig:CNOT2}c; the most apparent distinction is the generally larger control amplitudes for CNOT(ii). However, as was shown in the NH analysis of CNOT(i), the reduced field amplitudes did not prevent the creation of suitably interfering pathway amplitudes.

\subsection{SWAP Gate} \label{ssec:swap_gate}

\begin{figure}
    \includegraphics[width=\linewidth]{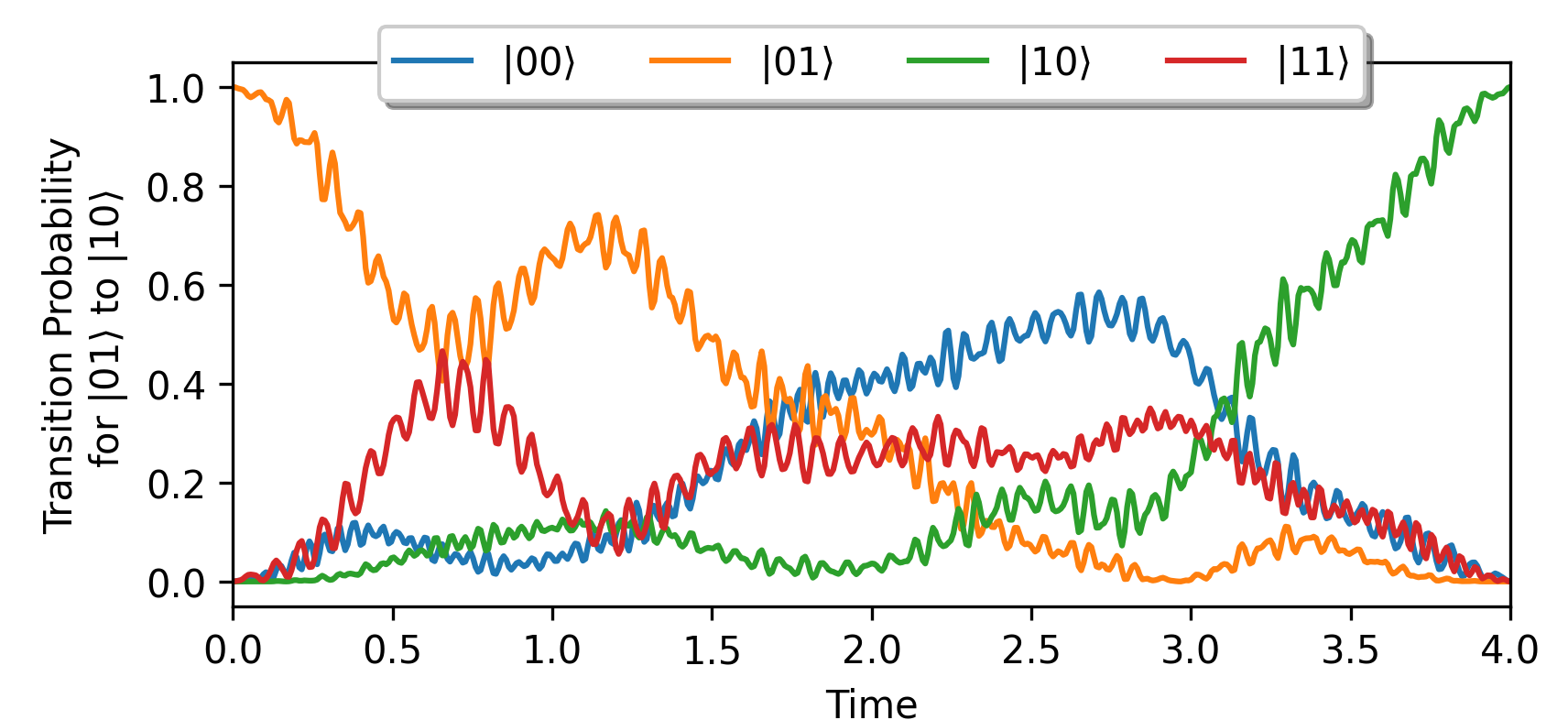}
    \caption{\justifying Populations of all four computational basis states driven by the Hamiltonian in Eqs.~\eqref{eq:2q_h0} and \eqref{eq:2q_hc} under the effect of the SWAP gate acting on the initial state $\ket{01}$. }
    \label{fig:SWAPgate_population}
\end{figure}

The Hamiltonian for the SWAP gate is the same as the previous CNOT gates described by Eqs.~\eqref{eq:2q_h0} and \eqref{eq:2q_hc}. The only difference is that the control fields $\epsilon_x(t)$ and $\epsilon_y(t)$ are chosen to implement a SWAP gate
\begin{equation}
    \text{SWAP} = \mqty(1 & 0 & 0 & 0\\
    0 & 0 & 1 & 0\\
    0 & 1 & 0 & 0\\
    0 & 0 & 0 & 1).
\end{equation}
Similar to CNOT gates, the optimal control code used to generate the SWAP gate pulses added a physically irrelevant $45^\circ$ global phase that will appear in the mechanism analysis. The encoding and computation of pathways are done in the same manner as in the CNOT gates as well. This section will specifically focus on the system starting in the $\ket{01}$ state and the SWAP gate bringing it to $\ket{10}$, i.e., we are examining the mechanism for creating the $\mel{10}{U(T)}{01}$ matrix element of unit value and at chosen final time $T=4$. Notice that the control Hamiltonian in Eq.~\eqref{eq:2q_hc} has no direct transition from $\ket{01}$ to $\ket{10}$, i.e. $\mel{10}{H_c(t)}{01} = 0$. Therefore the simplest pathways connecting the states $\ket{01}$ and $\ket{10}$ will be  $01 \tto 00 \tto 10$ and $01 \tto 11 \tto 10$ correspondingly using $\ket{00}$ and $\ket{11}$ as intermediate states. The use of these pathways is reflected in the population plots in Figure 
\ref{fig:SWAPgate_population}, where all four states are present during the evolution.

\begin{figure}
    \includegraphics[width=\linewidth]{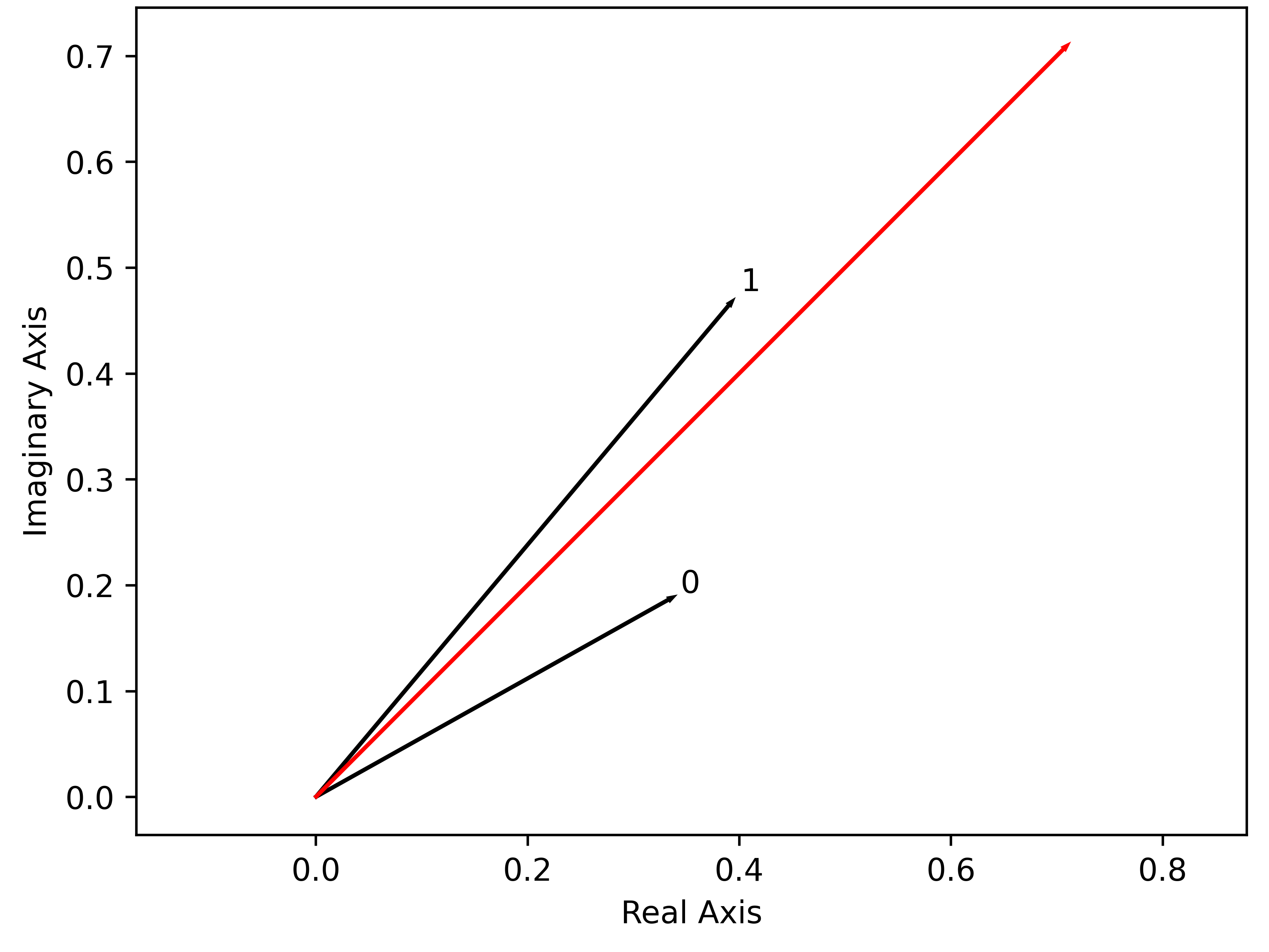}
    \caption{\justifying The Hermitian pathway class amplitudes of the SWAP gate for pathways corresponding to $\ket{01} \to \ket{10}$. Each amplitude is drawn as a vector and labeled with its associated frequency in Table \ref{tab:SWAP_pathways_01_10}. The red vector is the sum of all pathway class amplitudes and has a magnitude equal to 1. These Hermitian pathway classes are expanded upon in Table \ref{tab:SWAP_pathways_01_10}.}
    \label{fig:SWAPgate_arrow_plot_01_10}
\end{figure}

Mechanism analysis will be used to assess predictions inferred above from the population plots. This analysis is an ideal case for considering Hermitian pathway classes. In particular, an interesting question is whether the simplest possible pathways $01 \tto 00 \tto 10$ and $01 \tto 11 \tto 10$ are being used disregarding potential backtracking or Rabi flopping. Hermitian encoding was performed to reveal the associated pathway class amplitudes which are listed in Table \ref{tab:SWAP_pathways_01_10}. As anticipated from Figure~\ref{fig:SWAPgate_population}, the two most significant pathway classes are $[01 \tto 00 \tto 10]^\text{H}$ and $[01 \tto 11 \tto 10]^\text{H}$. These two pathways are depicted as complex vectors in Figure \ref{fig:SWAPgate_arrow_plot_01_10}. There were other Hermitian pathway classes but their magnitudes are comparably small and insignificant to the overall mechanism as evident in Table \ref{tab:SWAP_pathways_01_10}.

\begin{table*}
\centering
\begin{tabular}{c l c c} 
 \hline
 \hline
 \addlinespace[2pt]$\gamma^{n(l_1,\dots,l_{n-1})}_{ba}$ & Hermitian Pathway Class & Magnitude & Phase \\ [0.5ex] 
 \hline
 \addlinespace[1pt] 1   & $[01 \tto 00 \tto 10]^\text{H}$ & 0.612 & 50$^\circ$   \\
\addlinespace[1pt] 0    & $[01 \tto 11 \tto 10]^\text{H}$ & 0.387 & 29$^\circ$   \\
\addlinespace[1pt] -1   & $[01 \tto 11 \tto 10 \tto 00 \tto 01 \tto 11 \tto 10]^\text{H}$ & 0.053 & 140.$^\circ$  \\
\addlinespace[1pt] 2    & $[01 \tto 00 \tto 10 \tto 11 \tto 01 \tto 00 \tto 10]^\text{H}$ & 0.022 & 45$^\circ$  \\
 $\cdots$ & & $\cdots$ & $\cdots$ \\
 \hline
 Sum &  & 1.000 & 45$^\circ$ \\ 
 \hline
 \hline
\end{tabular}
\caption{\justifying Frequencies, Hermitian pathway classes, amplitude magnitudes, and amplitude phases of the SWAP gate for the matrix element $\mel{10}{U(T)}{01}$. All pathway classes with a magnitude greater than $0.01$ are listed. }
\label{tab:SWAP_pathways_01_10}
\end{table*}

\begin{figure}
    \includegraphics[width= \linewidth]{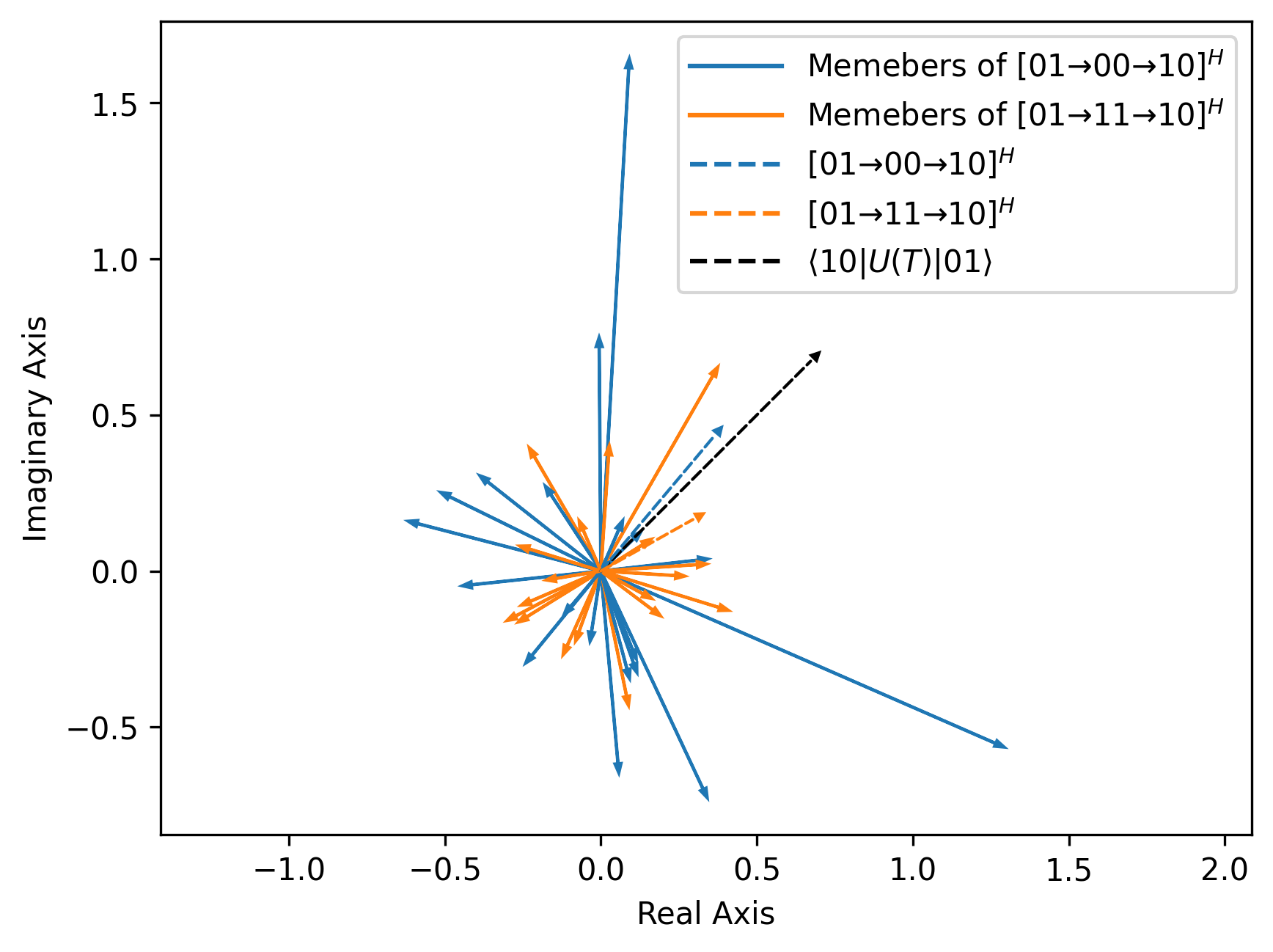}
    \caption{\justifying The non-Hermitian pathway class amplitudes of the SWAP gate for pathways corresponding to $\ket{01} \to \ket{10}$. The solid blue color indicates members of the $[01 \tto 00 \tto 10]^H$ pathway class, while the solid red color indicates members of the $[01 \tto 11 \tto 10]^H$ pathway class. The dashed blue and red pathways are the collective pathway amplitudes corresponding to $[01 \tto 00 \tto 10]^H$ and $[01 \tto 11 \tto 10]^H$, respectively. The dashed black arrow is the sum of all pathway class amplitudes and has a magnitude equal to 1, which includes interference between  $[01 \tto 00 \tto 10]^H$ and $[01 \tto 11 \tto 10]^H$. }
    \label{fig:SWAPgate_arrow_plot_comparison}
\end{figure}

An interesting mechanistic detail to explain is why the $[01 \tto 00 \tto 10]^H$ pathway class {has} a 20\% larger magnitude than the  $[01 \tto 11 \tto 10]^H$ pathway class. This issue can be answered by examining the mechanism in more depth using NH pathway classes that include backtracking information. In Figure \ref{fig:SWAPgate_arrow_plot_comparison}, the results of an NH Hamiltonian encoding are presented as a complex vector plot. The NH pathways classes belonging to $[01 \tto 00 \tto 10]^H$ are colored in blue while NH pathways classes belonging to $[01 \tto 11 \tto 10]^H$ are colored in red. These colors make it easy to see the constructive and destructive interferences within and between the pathway classes. Many of the most significant non-Hermitian pathways belong to the $[01 \tto 00 \tto 10]^H$ pathway class. Of the seven largest pathways, six belong to $[01 \tto 00 \tto 10]^H$. The presence of large pathways within the $[01 \tto 00 \tto 10]^H$ means that the many other blue pathway classes with smaller magnitudes likely do not play as important roles in the mechanism. This is in contrast to the $[01 \tto 11 \tto 10]^H$ pathway class where there is only one red pathway that is significantly larger than the rest. This result implies that the destructive interference between the many small-magnitude pathways plays a significant role in the mechanism for the $[01 \tto 11 \tto 10]^H$ pathway class. This collective behavior reveals how NH encoding can explain subtle mechanistic features like the different roles of $\ket{00}$ and $\ket{11}$ as intermediate states. Once again, a different optimal control field could have its own mechanism to assess for the SWAP gate, but the same family of Hamiltonian encoding tools may be applied. 

\section{Conclusion} \label{sec:conclusion}

The determination of quantum pathways through Hamiltonian encoding provides a means for obtaining a better understanding of the mechanism behind quantum dynamical operations. The original work has been applied to many situations, and this paper serves to use it for one and two-qubit gates that are common in the quantum information sciences. Although the Hamiltonians utilized in the paper are commonly employed, the same mechanism analysis techniques can also be used on alternative qubit models with different physical implementations. In this paper, control fields for an X-gate, two CNOT gates, and a SWAP gate were investigated and interesting properties of the gate creation mechanisms were explained using quantum pathway amplitudes through Hamiltonian encoding. Prevailing themes in the analysis were (a) the importance of constructive/destructive interference and (b) information revealed by encoding the dynamics that would otherwise likely be hidden from normal state population observations.

The concepts and computational tools set out in this paper may be applied in several potential future investigations. For example, these techniques can be applied to appropriate robust and non-robust control formulations of noisy quantum systems with either uncertainty in the control or lack of full knowledge of the Hamiltonian. The revealed quantum pathways may help better understand how uncertainty affects the fidelity of a controlled gate. With this knowledge, it is also possible that mechanism analysis can be used in the design of optimal controls to mitigate the effects of uncertainty. In the same vein, mechanism analysis can be used to study open quantum systems to explore the effect of the environment. This situation requires a redefinition of the pathways to either use the Lindbladian as was done in \cite{PhysRevA.93.053407} or another suitable form of open system model. Finally, the analysis performed in the paper was only on single and two-qubit systems while a practical quantum computer needs many qubits and pathway analysis can be performed on these larger systems. However, taking on such an expanded task becomes computationally expensive as the number of qubits increases. Future work will explore these additional directions.

\begin{acknowledgments}
M.K. acknowledges support from the Program in Plasma Science and Technology under U.S. Department of Energy (DOE) contract DE-AC02-09CH11466. H.R. and G.B. acknowledge support from the DOE under Grant DE-FG-02ER15344.
\end{acknowledgments}

\bibliography{sources}

\end{document}